\documentclass[prl, pdflatex, twocolumn, superscriptaddress, notitlepage, floatfix, footnotebib]{revtex4-2}
\usepackage{xcolor}
\usepackage{graphicx}
\usepackage{wrapfig}
\usepackage{amsmath}
\usepackage{float}
\usepackage{amssymb}
\usepackage{multirow}
\usepackage{slashed}
\usepackage{physics}
\usepackage{booktabs}
\usepackage[colorlinks=true, linkcolor=blue, citecolor=blue, urlcolor=blue]{hyperref}

\newcommand \ml {m_l}
\newcommand \ms {m_s}
\newcommand \U {\mathcal{U}}
\newcommand \av[2] {\left\langle{#1}\right\rangle_{#2}}
\newcommand \pbp {\bar{\psi}\psi}
\newcommand \cum[1] {\mathbb{K}_{#1}}
\newcommand \lda {\lambda}
\newcommand \ru[1] {\rho_U(\lda_{#1})}
\newcommand \pu[2] {P_U(\lda_{#1};{#2})}
\newcommand \pn {P_n(\lda)}
\newcommand \fn {f_n(z)}
\newcommand \gn {g_n(\lda)}
\newcommand \nt {N_\tau}
\newcommand \ns {N_\sigma}
\newcommand \tc {T_c}
\newcommand \hlda {\hat{\lda}}
\newcommand \hp[1] {\hat{P}_{#1}(\hlda)}
\newcommand \hml {\hat{m}_l}
\newcommand \hgn {\hat{g}_n(\hlda)}
\newcommand \D {\slashed{D}}
\begin{document}
\title{Microscopic Encoding of Macroscopic Universality: Scaling Properties of Dirac Eigenspectra near QCD Chiral Phase Transition}
\author{H.-T. Ding}
\affiliation{Key Laboratory of Quark and Lepton Physics (MOE) and Institute of Particle Physics, Central China Normal University, Wuhan 430079, China}
\author{W.-P. Huang}
\email{huangweiping@mails.ccnu.edu.cn}
\affiliation{Key Laboratory of Quark and Lepton Physics (MOE) and Institute of Particle Physics, Central China Normal University, Wuhan 430079, China}
\author{Swagato Mukherjee}
\affiliation{Physics Department, Brookhaven National Laboratory, Upton, New York 11973, USA}
\author{P. Petreczky}
\affiliation{Physics Department, Brookhaven National Laboratory, Upton, New York 11973, USA}
\begin{abstract}
Macroscopic properties of the strong interaction near its chiral phase transition exhibit scaling behaviors, which are the same as those observed close to the magnetic transition in a 3-dimensional classical spin system with $O(4)$ symmetry. We show that the universal scaling properties of the chiral phase transition in quantum chromodynamics (QCD) at the macroscale are, in fact, encoded within the microscopic energy levels of its fundamental constituents, the quarks. We establish a connection between the cumulants of the chiral order parameter, i.e., the chiral condensate, and the correlations among the energy levels of quarks, i.e., the eigenspectra of the massless QCD Dirac operator. This relation elucidates how the fluctuations of the chiral condensate arise from the correlations within the infrared part of the energy spectra of quarks, and naturally leads to a generalization of the Banks-Casher relation for the cumulants of the chiral condensate. Then, through (2+1)-flavor lattice QCD calculations with varying light quark masses near the QCD chiral transition, we demonstrate that the correlations among the infrared part of the Dirac eigenvalue spectra exhibit same universal scaling behaviors as expected of the cumulants of the chiral condensate. We find that these universal scaling behaviors extend up to the physical values of the up and down quark masses. Our study reveals how the hidden scaling features at the microscale give rise to the macroscopic universal properties of QCD.
\end{abstract}
%
% \date{\today}
%
\maketitle
\emph{Introduction.---}
Critical phenomena exhibited in the vicinity of continuous second order phase transition are ubiquitous in nature~\cite{ma2000modern}. Phase transitions at high temperatures involving the electroweak and strong forces of nature have given rise to the Universe as we experience today. Lattice-regularized field theory calculations have shown that the high temperature transitions in matters governed by the electroweak~(see, e.g., discussions in~\cite{DOnofrio:2014rug}) and strong interactions~\cite{Aoki:2006we,Bhattacharya:2014ara,HotQCD:2018pds,Borsanyi:2020fev} are rapid crossovers. But for small enough Higgs mass ~\cite{Kajantie:1995kf,Kajantie:1996mn,Karsch:1996yh,Kajantie:1996qd,Gurtler:1997hr,Csikor:1998eu} and in the massless (chiral) limit of up and down quarks~\cite{Pisarski:1983ms} the electroweak and strong forces, respectively, are expected to
undergo true phase transitions. 

In the vicinity of a second order phase transition macroscopic quantities related to the order parameter exhibit telltale scaling behaviors that are uniquely characterized by the dimensionality and global symmetries of the system, irrespective of the details of its microscopic degrees of freedom and interactions. Based on the global symmetries of quantum chromodynamics (QCD), the theory of strong interaction, the second order QCD chiral transition can be in the 3-dimensional $O(4)$ universality class~\cite{Pisarski:1983ms}~\footnote{It is also possible that the second order chiral phase transition can be in a different universality class from the 3-dimensional $O(4)$ if the $U_A(1)$ anomaly is largely weakened at the chiral phase transition temperature~\cite{Butti:2003nu, Pelissetto:2013hqa, Grahl:2014fna}. Recent lattice QCD results extrapolated to the chiral and continuum limits have shown that the 3-dimensional $O(4)$ universality class is preferred~\cite{Ding:2020xlj}.}. 
For lattice-regularized QCD using costly chiral fermions, e.g., overlap fermions, the chiral symmetry of QCD is strictly preserved for any finite value of the regulator, i.e., for nonvanishing lattice spacing, the universality class is 3-dimensional $O(4)$; while for the case using staggered fermions the chiral symmetry is only partially preserved and the universality class falls into 3-dimensional $O(2)$~\cite{Kilcup:1986dg,Ejiri:2009ac,Clarke:2020htu,Lahiri:2021lrk}. With present computational resources it is only feasible to carry out large scale lattice QCD simulations toward a chiral limit of up and down quarks using the staggered fermions. Thus, lattice QCD studies of chiral phase transition using staggered fermion discretizations are expected to observe the same macroscopic scaling behaviors as that in the vicinity of the liquid to superfluid $\lambda$ transition in $^4\mathrm{He}$~\cite{chaikin_lubensky_1995}; notwithstanding, the microscopic degrees of freedom for QCD are quarks and gluons governed by the strong force while for $^4\mathrm{He}$ are electrons and photons interacting via the electromagnetism. Because of this universal feature, to understand and predict macroscopic properties of a system close a second order phase transition one, most often, resorts to a simplified effective theory possessing the same dimensionality and global symmetries of the original theory, ignoring its microscopic complexities~\cite{chaikin_lubensky_1995}.

However, it is unclear if and how the macroscopic universal scaling properties of the strong interaction in the vicinity of the chiral phase transition are concealed within the microscopic energy spectrum of its elementary degrees of freedom, quarks. The goal of this work is lattice QCD-based understanding of possible connections between the universal features at the macroscopic and microscopic scales of QCD. An analogous goal for quantum electrodynamics will be to comprehend how the macroscopic scaling properties near the $\lambda$ transition of $^4\mathrm{He}$ arise from the energy levels of electrons without resorting to an effective theory. 

In this Letter, first, we will establish theoretical relations between the cumulants of the chiral condensate and correlations among the eigenvalue spectrum of the massless Dirac operator. We will then demonstrate how the $O(2)$ scaling properties of the cumulants of the chiral condensate are reflected within the correlations of the Dirac eigenvalue spectrum through the state-of-the-art lattice QCD calculations in the staggered discretization scheme.

\emph{Theoretical developments.---}
For the lattice QCD calculations we will use (2+1)-flavor QCD with degenerate light up ($u$) and down ($d)$ quarks having masses  $\ml=m_u=m_d$ and a heavier strange quark with physical mass $\ms$. Since the physical strange quark plays no significant role in the discussion of the $O(2)$ critical behavior, for simplicity in this subsection we develop the main theoretical idea by considering QCD with two degenerate light quarks.

Consider the Euclidean-time  QCD action $S[\U,\ml] = S_g[\U] + \bar{\psi}\D[\U]{\psi} + \ml\pbp$, where $\pbp = \bar\psi_u\psi_u + \bar\psi_d\psi_d$, $S_g[\U]$ is the  pure gauge action, and $\D[\U]$ is the massless QCD Dirac operator for given background $SU(3)$ gauge field $\U$. To probe the system in chiral limit, $\ml\to0$, we introduce a probe operator, $\pbp(\epsilon)\equiv 2\text{Tr}(\D[\U] + \epsilon)^{-1}$, with the background $\U$ distributed according to $\exp{-S[\U,0]}$. The valance quark mass, $\epsilon>0$, is introduced to facilitate the evaluation of the probe operator. Traces over the color, spin and space-time indices are denoted by $\mathrm{Tr}$. 

The $n$th order cumulants, $\cum{n}$, of the order parameter $\pbp(\ml)$ can be obtained from the generating functional
\begin{align}
  \mathbb{G}(\ml;\epsilon) = \ln \av{ \exp \left\{ - \ml \pbp(\epsilon) \right\} } {0} \,,
\label{eq:gf1}
\end{align}
as  
\begin{align}
  \cum{n} \left[ \pbp \right]= \frac{T}{V} \, (-1)^n \left. \frac{\partial^n \mathbb{G}(\ml;\epsilon)}{\partial \ml^n} \right\vert_{\epsilon=\ml} \,.
\label{eq:kn1}
\end{align}
Hereafter, $T$ is the temperature and $V$ is the spatial volume of the system, and $\av{\cdot}{0}$ denotes expectation value with respect to the QCD partition function in the chiral limit, $Z(0)=\int \exp{-S[\U,0]} \mathcal{D}[\U]$. With $\av{\cdot}{}$ the  expectation value with respect to the QCD partition function $Z(\ml)=\int \exp{-S[\U,\ml]} \mathcal{D}[\U]$, and recognizing $\av{\mathcal{O}}{} = \av{ \mathcal{O} \exp\{ - \ml \pbp(\ml) \} } {0}/\av{\exp\{ - \ml \pbp(\ml) \} } {0}$ and $Z(\ml)/Z(0)= \av{ \exp\{ - \ml \pbp(\ml) \} } {0}$, it is easy to see that $\cum{n}$ are the standard cumulants of $\pbp(\ml)$; e.g., $\cum{1}\left[ \pbp \right]=T\langle\pbp(\ml)\rangle/V$, $\cum{2}\left[ \pbp \right]=T\langle[\pbp(\ml)-\langle\pbp(\ml)\rangle]^2\rangle/V$, $\cum{3}\left[ \pbp \right]=T\langle[\pbp(\ml)-\langle\pbp(\ml)\rangle]^3\rangle/V$, etc.

Energy levels of a massless quark in the background of $\U$ are given by the eigenvalues, $\lda_j[\U]$, of $\D[\U]$. In terms of $\lda_j[\U]$ the probe operator can be expressed as $\pbp(\epsilon)\equiv 2\text{Tr}(\D[\U] + \epsilon)^{-1} = 2 \sum_j  (i\lda_j+\epsilon)^{-1}$. Thus, \autoref{eq:gf1} becomes
\begin{align}
  \mathbb{G}(\ml;\epsilon) = \ln \av{ \exp \left\{ - \ml \int_0^\infty \negthickspace \negthickspace \pu{}{\epsilon} d\lda  \right\} } {0} \,,
\label{eq:gf2}
\end{align}
where 
\begin{align}
  \pu{}{\epsilon} = \frac{4\epsilon\ru{}}{\lda^2 + \epsilon^2} \,, 
  ~ \text{and} ~
  \ru{} = \sum_j \delta(\lda-\lda_j) \,.  
\label{eq:pu}
\end{align}
From \autoref{eq:kn1} it is straightforward to obtain
\begin{align}
  \cum{n} [\pbp] = \int_0^\infty \negthickspace \negthickspace \pn d\lda  \,,
\label{eq:kn2}
\end{align}
where $P_1(\lda)=K_1[\pu{}{\ml}]$ for $n=1$, and for $n\geq2$
\begin{align}
  P_n(\lda)=\int_0^\infty & K_1\big[\pu{}{\ml}, \pu{2}{\ml}, \dotsc, \pu{n}{\ml} \big] \nonumber \\
                & \times\prod_{i=2}^{n}d\lda_i  \,.
\label{eq:pn}
\end{align}
Here, $K_1$ is the first order joint cumulant of $n$ variables ($X_i$) defined as 
\begin{align}
  K_1 (X_1,\cdots,X_n)= \frac{T}{V} \, (-1)^n \left. \frac{\partial^n \ln\av{\prod_{i=1}^{n} e^{-t_i X_i}}{} }{\partial t_1 \dotsm \partial t_n}  \right\vert_{t_1, \dotsc,t_n=0} \,.
\label{eq:K1}
\end{align}

\autoref{eq:kn2} is our main theoretical result connecting the cumulants of the order parameter to the $n$-point correlations of the quark energy levels $\ru{}$. The explicit expressions of $\cum{1}[\pbp]$, $\cum{2}[\pbp]$, and $\cum{3}[\pbp]$ in terms of $\ru{}$ are provided in \autoref{eq:eq:Kn-rhoU} of Supplemental Material.

The chiral phase transition in the staggered lattice QCD at nonvanishing lattice spacings is expected to be in the 3-dimensional $O(2)$ universality class. Following the expectations for a 3-dimensional $O(2)$ spin model near criticality~\cite{Engels:2001bq}, in the vicinity of the chiral transition
\begin{align}
  \cum{n}[\pbp] = \int_0^\infty \negthickspace \negthickspace \pn d\lda \sim \ml^{1/\delta-n+1} f_n(z) \,.
\label{eq:scaling}
\end{align}
The scaling variable $z \propto z_0 \ml^{-1/\beta\delta}(T-\tc)/\tc$, where $\tc$ is the chiral phase transition temperature and $z_0$ is a scale parameter; both are system specific. $\beta$ and $\delta$ are the universal critical exponents, and $f_{n+1}(z) = (1/\delta-n+1) f_{n}(z) - z f_{n}^{\prime}(z) / {\beta\delta}$ are the universal scaling functions of 3-dimensional $O(2)$ universality class. Here, $n\ge1$ and the superscript prime denotes derivative with respect to $z$~\footnote{According to the notation in the literature $f_1(z)\equiv f_G(z)$ and $f_2(z)\equiv f_\chi(z)$.}. In our work we adopted $\beta=0.349$, $\delta=4.78$ and for consistency the scaling functions $f_n(z)$ of the $O(2)$ universality class determined from Refs.~\cite{Engels:2001bq,Ejiri:2009ac}. 

\autoref{eq:scaling} indicates that universal scaling properties of the macroscopic observables $\cum{n}[\pbp]$ arise from the correlations among the microscopic energy levels $\pn$. To elucidate this point, consider $\ml\to0$. Then from \autoref{eq:pu} one finds $\pu{}{\epsilon\to0} = 2\pi \ru{} \delta(\lda)$, giving $\lim_{\ml\to0}P_1(\lambda)=2\pi K_1[\ru{}]\delta(\lda)$ and $\lim_{\ml\to0} \pn = (2\pi)^n K_1[\ru{}, [\rho_U(0)]^{n-1}] \delta(\lda)$ for $n\geq2$ (from \autoref{eq:pn}). Noting that $K_1$ of $n$ identical variables is equivalent to $\cum{n}$, in the chiral limit \autoref{eq:kn2} thus becomes a generalization of the Banks-Casher relation~\cite{Banks:1979yr} expressed as follows:
\begin{align}
  \lim_{\ml\to0} \cum{n} [\pbp] = (2\pi)^n \cum{n} [\rho_U (0)] \,.
\label{eq:genBC}
\end{align}
To the best of our knowledge this generalized relation between the higher order cumulants of chiral condensate in the chiral limit and density of the deep infrared energies of quarks is new in literature. 

\begin{figure*}[!htp]
  \centering
    \includegraphics[width=0.32\textwidth,height=0.17\textheight]{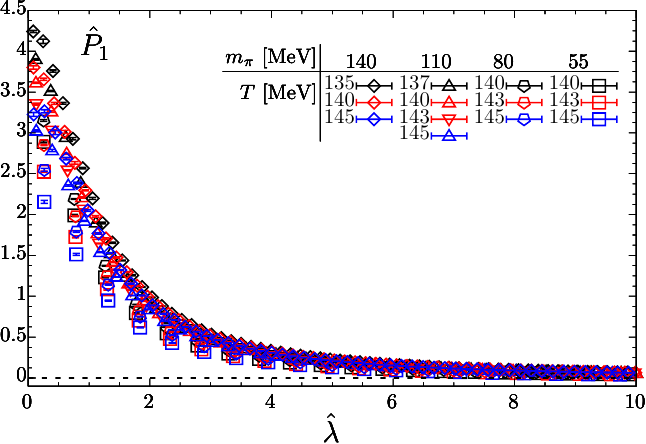}
    \includegraphics[width=0.32\textwidth,height=0.17\textheight]{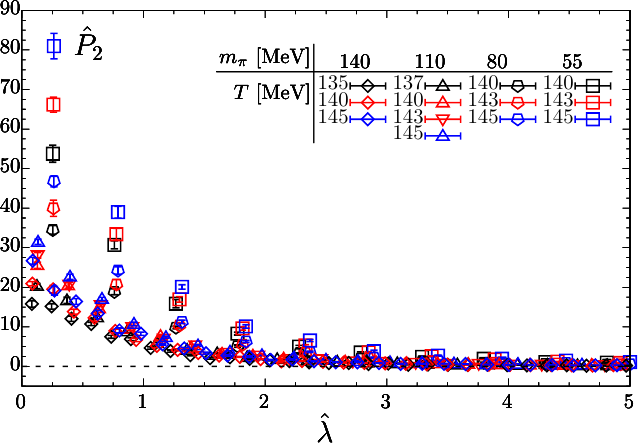}
    \includegraphics[width=0.32\textwidth,height=0.17\textheight]{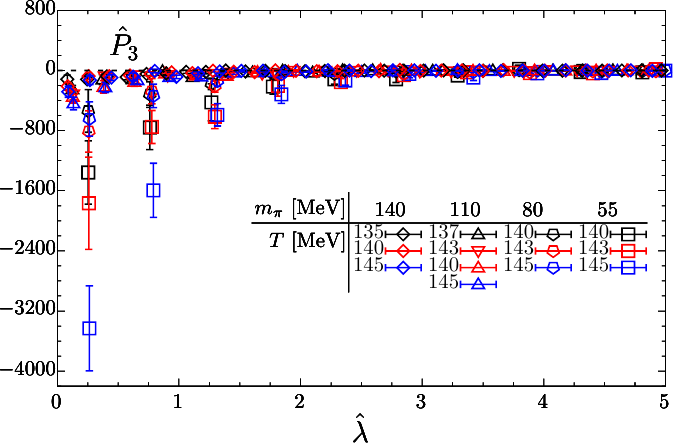}
  	\caption{$\hp{1}$ (left), $\hp{2}$ (middle) and $\hp{3}$ (right) for 135~MeV~$\le T \le$~145~MeV and 55~MeV~$\le m_\pi \le$~ 140~MeV.}
	\label{fig:P123}
\end{figure*}

In the chiral limit  and close to $\tc$, $\cum{n}[\pbp]$ should manifest universal scaling, e.g., $\cum{1}[\pbp] \sim \abs{{(T-T_c)}/{T_c}}^{\beta}$ and $\cum{2}[\pbp] \sim \abs{{(T-T_c)}/{T_c}}^{\beta(1-\delta)}$. According to \autoref{eq:genBC} this must arise from the universal behaviors of the $\lambda$-independent $\cum{n}[\rho_U(0)]$. Thus, it is natural to expect that for small $\ml$ within the scaling window the critical scaling of $\cum{n}[\pbp]$ in \autoref{eq:scaling} arises from the universal behaviors of the amplitudes of $\pn$ at the infrared, and not from its system-specific $\lambda$ dependence; i.e.,
\begin{align}
  \pn = \ml^{1/\delta-n+1} \fn \gn \,. 
\label{eq:pn1}
\end{align}
Here, $\gn$ are nonuniversal functions encoding the properties of the specific system under consideration. 

Next, we numerically establish \autoref{eq:pn1} through lattice QCD calculations. 

\emph{Lattice QCD calculations.---}
Lattice QCD calculations were carried out between $T=$~$135-176$~MeV for (2+1)-flavor QCD using the highly improved staggered quarks and the tree-level Symanzik gauge action, a setup extensively used by the HotQCD Collaboration~\cite{Bazavov:2010ru,Bazavov:2011nk, Bazavov:2014pvz, Bazavov:2012jq, Bazavov:2017dus, Bazavov:2018mes}. $\ms$ was fixed to its physical value with a varying $\ml = \ms/27, \ms/40, \ms/80, \ms/160$, which correspond to the Goldstone pion masses $m_{\pi}\approx 140, 110, 80, 55$~MeV, respectively. The temporal extents of the lattices were $\nt=8$, and spatial extents were chosen to be $\ns=(4-7)\nt$. The gauge field configurations were generated using a software suite \texttt{SIMULATeQCD}~\cite{Bollweg:2021cvl},  and the same gauge ensembles were used for the determination of chiral phase transition temperature in the continuum limit~\cite{HotQCD:2019xnw}. 

Observables were calculated on gauge configurations from every tenth molecular dynamics trajectory of unit length, after skipping at least first $800$ trajectories for thermalization. $\ru{}$ and $\pn$ for $n = 1, 2, 3$ over the entire range of $\lda$ were computed using the Chebyshev filtering technique combined with the stochastic estimate method~\cite{Ding:2020eql, Ding:2020xlj, Ding:2021gdy, Giusti:2008vb, Cossu:2016eqs, Fodor:2016hke} on about $3000$ configurations. Orders of the Chebyshev polynomials were chosen to be $2\times10^5$ and $24-96$ Gaussian stochastic sources were used. Exact details are provided in \autoref{sup_table_setup} of Supplemental Material. 

\emph{Results.---}
Owing to \autoref{eq:genBC} we expect the relevant infrared energy scale is $\lda\sim\ml$ for small values of $\ml$. It is natural to express all quantities as functions of the dimensionless and renormalization group invariant $\lda/\ml$: 
\begin{align}
\begin{split}
  & \hlda = \lda / \ml \,, \quad 
  \hml = \ml / \ms \,, z=z_0\hml^{-1/\beta\delta}(T-T_c)/T_c\,,\\
  & \hp{n} = \ms^{n+1} \hml P_n(\lda) / \tc^4 \,, \quad \text{and}   \\
  & \mathbb{\hat K}_{n} [\pbp] = \int_0^\infty \negthickspace \negthickspace \hp{n} d\hlda \sim \hml^{1/\delta-n+1} f_n(z) \,,
\end{split}
\label{eq:hatdef}
\end{align}
where the dimensionless and renormalization group invariant $\mathbb{\hat K}_{n} [\pbp] = \ms^n \cum{n} [\pbp] / \tc^4$. 

In~\autoref{fig:P123} we show $\hp{n}$ for $n=1$, 2, 3 as a function of $\hlda$ in the proximity of $\tc=144.2(6)$~MeV~\cite{Clarke:2020htu}. $\hp{n}$ rapidly vanishes for $\hlda\gtrsim1$, and the regions where $\hp{n}\ne0$ get smaller with increasing $n$. This reinforces that the relevant infrared energy scale turns out to be $\hlda\sim1$. In this infrared region $\hp{n}$ at a fixed $T$ shows clear dependences on $\ml$, which becomes stronger for increasing $n$. The form of $\ml$ dependence of $\hp{n}$ also changes with varying $T$. 

As shown in \autoref{fig:K123-reproduce} of Supplemental Material we have checked that integrals over the relevant nonvanishing infrared regions of $\hp{n}$ (cf. \autoref{eq:hatdef}) reproduces $\cum{n}[\pbp]$, independently calculated through inversions of the fermion matrices, for $n=1,2,3$. As seen from \autoref{fig:P123}, expectedly, our results become increasingly noisy with increasing $n$ and decreasing $\ml$. With our present statistics we cannot access correlation functions $n>3$, particularly for smaller $\ml$.   

\begin{figure*}[!htp]
  \centering
    \includegraphics[width=0.32\textwidth,height=0.17\textheight]{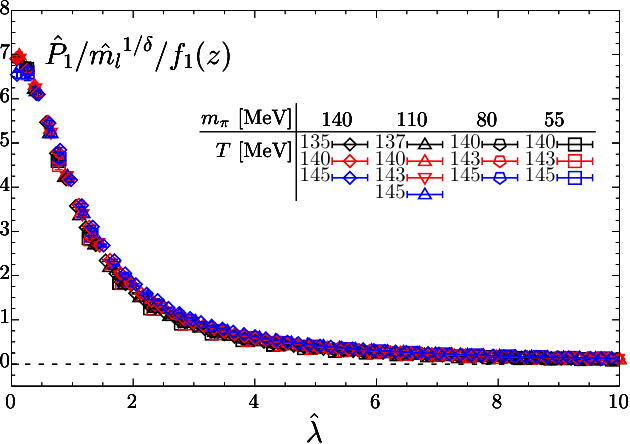}
    \includegraphics[width=0.32\textwidth,height=0.17\textheight]{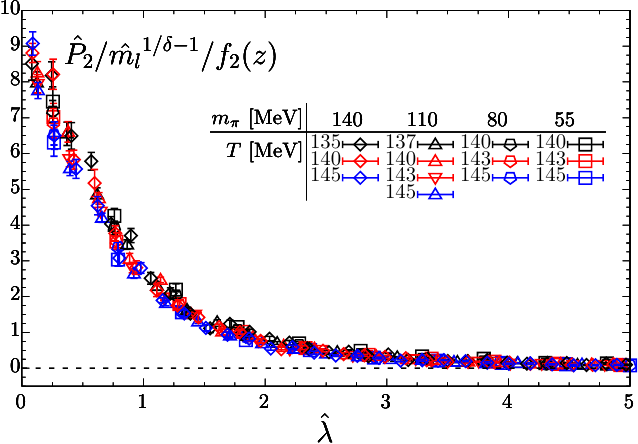}
    \includegraphics[width=0.32\textwidth,height=0.17\textheight]{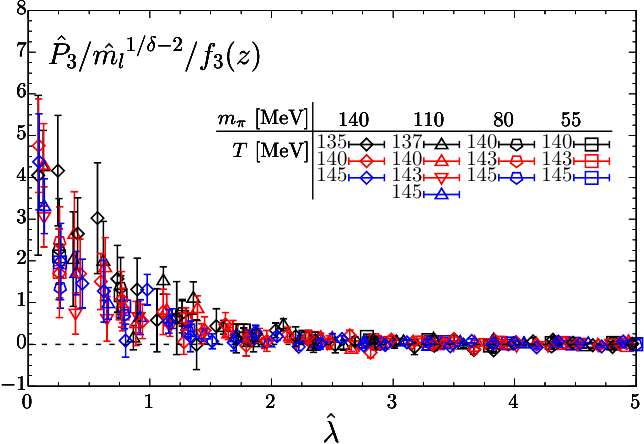}
  	\caption{$\hp{n}$ in \autoref{fig:P123} rescaled by $\hml^{1/\delta+1-n}f_n(z)$ for $n=1$ (left), $n=2$ (middle) and $n=3$ (right).}
 \label{fig:RescaledP123}
\end{figure*}

The $\ml$ and $T$ dependence of $\hp{n}$ shown in \autoref{fig:P123} can be understood in terms of the 3-dimensional $O(2)$ scaling properties. Once the $\hp{n}$ are rescaled with respective $\hml^{1/\delta+1-n}f_n(z)$ the data in \autoref{fig:P123} magically collapse onto each other, see \autoref{fig:RescaledP123}. The system-specific parameters $\tc=144.2(6)$ MeV and $z_0=1.83(9)$ needed to obtain $f_n(z)$ were taken from Ref.~\cite{Clarke:2020htu}, where 3-dimensional $O(2)$ scaling fits were carried out for the same lattice ensembles but using an entirely different macroscopic observable, namely the $\ml$ dependence of the static quark free energy. Thus, our expectations from \autoref{eq:pn1} are clearly borne out in \autoref{fig:RescaledP123}, namely 
\begin{align}
  \hp{n} = \hml^{1/\delta-n+1} \fn \hgn \,,
\label{eq:pn2}
\end{align}
where $\hgn$ characterize the system specific of the $n$th order energy-level correlations. To satisfy our generalized Banks-Casher relations of \autoref{eq:genBC} the $\hgn$ must also satisfy $\lim_{V\to\infty} \lim_{a\to0} \lim_{\ml\to0} \hgn \to \delta(\hlda)$, such that $\cum{n}[\pbp]$ has the correct scaling behavior in $(T-T_c)/T_c$.  

The values of $z_0$ and $\tc$ used to demonstrate the universal scaling in \autoref{fig:RescaledP123} were obtained fitting lattice results for $\ml$ dependence of the static quark free energy only for 55~MeV~$\le m_\pi \le$~110~MeV~\cite{Clarke:2020htu}. It is noteworthy that the physical QCD with $m_\pi\approx$~140~MeV also shows the same universal scaling for 135~MeV~$\le T \le$~145~MeV. Outside this temperature window we do not observe scaling (see \autoref{fig:RescaledP123-noscaling} of Supplemental Material).

As mentioned in Ref.~\cite{Clarke:2020htu}, presently $\{\tc,z_0\}$ are not very well determined. By using other values for $\{\tc,z_0\}$ quoted in Ref.~\cite{Clarke:2020htu} we checked that the scaling of \autoref{fig:RescaledP123} is fairly insensitive to the exact values of $\{\tc,z_0\}$ (see \autoref{fig:ReacaledP123-z0=2.24-Tc=145.6} of Supplemental Material). Presumably, this is because $\hp{n}$ are sensitive only to the deep infrared physics $\lda\sim\ml$. This is in contrast to many other macroscopic operators used for detailed scaling studies that contain large contributions from the ultraviolet energies~\cite{HotQCD:2019xnw,Ejiri:2009ac,Hegde:2015tbn,Dini:2021hug,Kaczmarek:2021ufg,Kotov:2021rah}. This suggests that it even might be advantageous to use $\hp{n}$ for detailed scaling studies to determine the system-specific parameters. Our focus here is to reveal the underlying connection between the universal features and the quark spectra, and such detailed scaling studies are beyond the scope of the present work.

\emph{Conclusions.---}
In this Letter, we investigate how the universal critical scaling of macroscopic observables near the QCD chiral transition arises from the microscopic degrees of freedom. We have presented a theoretical connection between the $n$th order cumulant of the chiral order parameter and the $n$-point correlations of the quark energy spectra. This connection led us to a generalized Banks-Casher relation, equating the $n$th order cumulant of chiral condensate to the $n$th order cumulant of the zero mode of the quark energy in the chiral limit. These new theoretical developments establish a direct connection between the universal scaling observed at the macroscale and the microscopic energy levels of the system. Through staggered lattice QCD calculations in the vicinity of the chiral phase transition with a series of light quark masses we have discovered the hidden universality within the  correlations among the quark energy spectra. We have found that these universal behaviors are also imprinted within the microscopic energy levels of QCD with physical light quark masses. 

The new theoretical developments presented in this work can be applied to spin systems near criticality. Away from the phase transition and at temperature much lower compared to $T_c$ with nonvanishing chiral condensate, the newly proposed quantity $P_n(\lambda)$ as well as the generalized Banks-Casher relation could be interesting to be investigated in the random matrix theory~\cite{Gasser:1987ah,Leutwyler:1992yt,Damgaard:2000gh,Damgaard:1997cy,Verbaarschot:2000dy}. The numerical techniques used here can be straightforwardly carried over to lattice QCD calculations with controlled thermodynamic and continuum limits when sufficient computing power is available. 

% \emph{Acknowledgements.---}
%
We thank Yu Zhang for the early involvement of the work, Sheng-Tai Li for technical support, Jacobus Verbaarschot for discussions on the random matrix theory, and the members of the HotQCD Collaboration for numerous interesting discussions. 
This material is based upon work supported partly by the National Key Research and Development Program of China under Contract No. 2022YFA1604900; the National Natural Science Foundation of China under Grants No. 12293064 , No. 12293060 and No. 12325508, and by the U.S. Department of Energy, Office of Science, Office of Nuclear Physics through Contract No. DE-SC0012704 and within the framework of Scientific Discovery through Advance Computing (SciDAC) award Fundamental Nuclear Physics at the Exascale and Beyond.
Numerical simulations have been performed on the GPU cluster in the Nuclear Science Computing Center at Central China Normal University (NSC$^3$), Wuhan supercomputing center, the facilities of the USQCD Collaboration, which are funded by the Office of Science of the U.S. Department of Energy. 
The gauge configurations used in this work were generated using computing resources obtained through PRACE grants at CSCS, Switzerland, and at CINECA, Italy as well as grants at the Gauss Centre for Supercomputing and NICJ$\ddot{\mathrm{u}}$lich, Germany. These grants provided access to resources on Piz Daint at CSCS and Marconi at CINECA, as well as on JUQUEEN and JUWELS at NIC. Additional calculations have been performed on GPU clusters of USQCD, at Bielefeld University, the PC2 Paderborn University, and NSC$^3$.

\bibliographystyle{apsrev4-1.bst}
\bibliography{ref.bib}
%
%%%%%%%%%%%%%%%%%%%%%%%%%%%%%%%%%%%%%
%%%%%%%%%%%%%%%%%%%%%%%%%%%%%%%%%%%%%
\newpage
\begin{widetext}

\section{Supplemental Materials}

We provide supplemental materials in the sequence according to the contents in the main material.

\section{I. $\cum{n} \left[ \pbp \right]$ in terms of $\ru{}$ and inversions of fermion matrix} 

The light quark chiral condensate $\pbp(\ml)$ can be expressed in terms of inversion of fermion matrix $\pbp(\ml)=2\text{Tr}(\D[\U] + \ml)^{-1}\equiv 2\text{Tr}~ M^{-1}$, where $M$ denotes the single-flavor light fermion matrix. Then $\cum{n}\left[ \pbp \right]$ can be directly measured through calculations of $\text{Tr}~ M^{-1}$ on the lattice, and for example $\cum{n} \left[ \pbp \right]$ for $n=1,2$ and 3 can be expressed in terms of $\text{Tr}~ M^{-1}$ as follows,
\begin{equation}
\begin{aligned}
    \cum{1}\left[ \pbp \right] & = \frac{2T}{V} \langle\operatorname{Tr} M^{-1}\rangle \,,
    \\
    \cum{2}\left[ \pbp \right] & = \frac{4T}{V} \left[\langle(\operatorname{Tr} M^{-1})^{2}\rangle-\langle\operatorname{Tr} M^{-1}\rangle^{2}\right] \,,
    \\
    \cum{3}\left[ \pbp \right] & = \frac{8T}{V}[\langle(\operatorname{Tr} M^{-1})^3\rangle-3\langle\operatorname{Tr} M^{-1}\rangle\langle(\operatorname{Tr} M^{-1})^2\rangle+2\langle\operatorname{Tr} M^{-1}\rangle^3 ] \,.
\end{aligned}
\label{eq:Kn-TrM}
\end{equation}
Note that $\cum{n} \left[ \pbp \right]$ is only related to $\operatorname{Tr} M^{-1}$ in direct measurements. Therefore no additional inversion of fermion matrix is needed to compute $\mathbb{K}_n$ for arbitrary $n$ on the lattice. 

The cumulants of the light quark chiral condensate $\cum{n}[\pbp]$ are explicitly expressed in terms of $\ru{}$ for $n \leq$ 3 as follows, 
\begin{align}
\begin{split}
%
   % \label{eq:K1-rhoU}
   \cum{1}[\pbp] &= \frac{T}{V} \av{\pbp (\ml)}{}
  = \int_0^\infty \negthickspace \negthickspace K_1[\pu{}{\ml}] ~\mathrm{d}\lda 
  = \frac{T}{V}\int_0^\infty \negthickspace \negthickspace \mathrm{d} \lda \frac { 4\ml \av{\ru{}}{} } {\lda^2+\ml^2} \,, 
   \\
  %\label{eq:K2-rhoU}
   \cum{2}[\pbp] &= \frac{T}{V}\av{[ \pbp(\ml) - \av{\pbp(\ml)}{} ]^2 }{} 
   =\int_0^\infty \negthickspace \negthickspace K_1[\pu{1}{\ml},\pu{2}{\ml}] ~\mathrm{d}\lda_1 \mathrm{d}\lda_2 \\
   &= \frac{T}{V} \int_0^\infty \negthickspace \negthickspace \mathrm{d}\lda_1 \mathrm{d}\lda_2 \frac{(4\ml)^2} {(\lda_1^2+\ml^2)(\lda_2^2+\ml^2)} \left[ \av{\ru{1} \ru{2}}{} - \av{\ru{1}}{} \av{\ru{2}}{} \right] \,, 
    \\
   %\label{eq:K3-rhoU}
   \cum{3}[\pbp] &= \frac{T}{V}\av{[ \pbp(\ml) - \av{\pbp(\ml)}{} ]^3 }{} 
   =\int_0^\infty \negthickspace \negthickspace K_1[\pu{1}{\ml},\pu{2}{\ml},\pu{3}{\ml}] ~\mathrm{d}\lda_1 \mathrm{d}\lda_2 \mathrm{d}\lda_3  \\
   &=\frac{T}{V} \int_0^\infty \negthickspace \negthickspace \mathrm{d}\lda_1 \mathrm{d}\lda_2 \mathrm{d}\lda_3 \frac{(4\ml)^3} {(\lda_1^2+\ml^2)(\lda_2^2+\ml^2)(\lda_3^2+\ml^2)} \bigg[\av{\ru{1} \ru{2} \ru{3}}{} - \av{\ru{1} \ru{2}}{}\av{\ru{3}}{} \\
   &
   - \av{\ru{1} \ru{3}}{}\av{\ru{2}}{} - \av{\ru{2} \ru{3}}{}\av{\ru{1}}{} + 2\av{\ru{1}}{}\av{\ru{2}}{}\av{\ru{3}}{}\bigg] \,.
\end{split}
\label{eq:eq:Kn-rhoU}
\end{align}

\section{II. Parameters and statistics of lattice QCD simulations}  %\label{sec:sup_setup}

In this subsection we list the simulation parameters as well as statistics in~\autoref{sup_table_setup}. 	

	\begin{table}[!htp]
	%\centering  	
	\begin{tabular}{*{7}{c}}
		\toprule \hline \hline
		\toprule  
	 	\multirow{3}*{$\beta$}  
	&	\multirow{3}*{$T$ [MeV]}	 	
	&   \multirow{3}*{$am_s$} 
	&   \multicolumn{4}{c}{$\#$ conf. [$N_{\mathrm{vec}}$ in Chebyshev]} 
	 \\  
		\cmidrule(r){4-7}
		\cline{4-7}
		& 
		& 
		& $m_{\pi}=140$ MeV
		& $m_{\pi}=110$ MeV
		& $m_{\pi}=80$ MeV
		& $m_{\pi}=55$ MeV
	\\
		\cmidrule(r){4-7}
		& 
		& 
		& ($N_{\sigma}^3\times N_{\tau}=32^3\times 8$)
		& ($N_{\sigma}^3\times N_{\tau}=40^3\times 8$)
		& ($N_{\sigma}^3\times N_{\tau}=56^3\times 8$)
		& ($N_{\sigma}^3\times N_{\tau}=56^3\times 8$)
	\\ 
	    \midrule \hline
		6.245 & 134.640 & 0.0830 & 4000 [78] & -- & -- & --
	    \\
	    6.260 & 136.792 & 0.0810 & -- & 4820 [96] & -- & --
	    \\
		6.285 & 140.449 & 0.0790 & 4580 [84] & 5020 [96] & 3060 [84] & 2840 [96]
		\\
		6.300 & 142.686 & 0.0772 & -- & 6570 [84] & 4000 [60] & 4660 [96]
		\\
		6.315 & 144.954 & 0.0760 & 4690 [84] & 5580 [84] & 4075 [60] & 3810 [96]
		\\
		6.330 & 147.254 & 0.0746 & -- & 5180 [84] & 3980 [48] & 3990 [72]
		\\
		6.354 & 151.001 & 0.0728 & 6830 [60] & 6000 [84] & 5000 [24] & --
		\\
		6.365 & 152.747 & 0.0716 & -- & 3240 [60] & -- & --		
		\\
		6.390 & 156.780 & 0.0694 & 6340 [60] & 4850 [48] & 2970 [48] & --
		\\
		6.423 & 162.246 & 0.0670 & 2820 [24] & 3150 [48] & 1380 [24] & --
		\\
		6.445 & 165.981 & 0.0652 & 3620 [24] & 2390 [24] & 1380 [24] & --
		\\
		6.474 & 171.017 & 0.0632 & 1600 [24] & 2800 [24] & -- & --
		\\
		6.500 & 175.642 & 0.0614 & 1450 [24] & 2920 [24] & -- & --
		\\
		\midrule
		\bottomrule \hline \hline		
	\end{tabular}
	\caption{Summary of lattice parameters; i.e., values of lattice gauge coupling $\beta$, temperature $T$, strange quark mass  $am_s$ in unit of lattice spacing, number of gauge configurations $\#$conf used in both the direct measurements of chiral observables and in the computation of $\rho_U$ via the Chebyshev polynomial method, as well as the number of random vectors $N_{\mathrm{vec}}$ used in the Chebyshev method shown in the square brackets. The order of Chebyshev polynomials are all set to $2\times10^5$. The pion mass $m_\pi$ and lattice size $N_{\sigma}^3\times N_{\tau}$ corresponding to different gauge ensembles are also listed.}
 \label{sup_table_setup}
\end{table}

\section{III. Supplementary Results}

\subsection{III.A. Reproduction of $\cum{n} \left[ \pbp \right]$ via $P_n(\lambda)$}

We show the temperature dependence of $\cum{n} \left[ \pbp \right]$ with $n\leq 3$ in \autoref{fig:K123-reproduce}. The bin-size of $\lambda$ needs to be fixed firstly when computing $\rho_U$, $P_n$ and the numerical integration of \autoref{eq:kn2}. In this work the bin-size of $\lambda$ was chosen such that the direct measurements of $\cum{2}\left[ \pbp \right]$ via fermion inversions (\textit{cf.} \autoref{eq:Kn-TrM}) can be reproduced via~\autoref{eq:eq:Kn-rhoU}. Similarly as in Ref.\cite{Ding:2020xlj, Ding:2021gdy}, once the bin-size of $\lambda$ is chosen to reproduce directly measured $\cum{2}\left[ \pbp \right]$, the same value can also be used to reproduce other directly measured $\cum{n} \left[ \pbp \right]$ without any further tuning. The consistency of $\cum{n} \left[ \pbp \right]$ for $n=1,2,3$ given by the two different methods as shown in \autoref{fig:K123-reproduce} verifies the reliability of the results of $P_n$ ($n=1,2,3$).

\begin{figure}[!htp]
  \centering
    \includegraphics[width=0.32\textwidth,height=0.17\textheight]{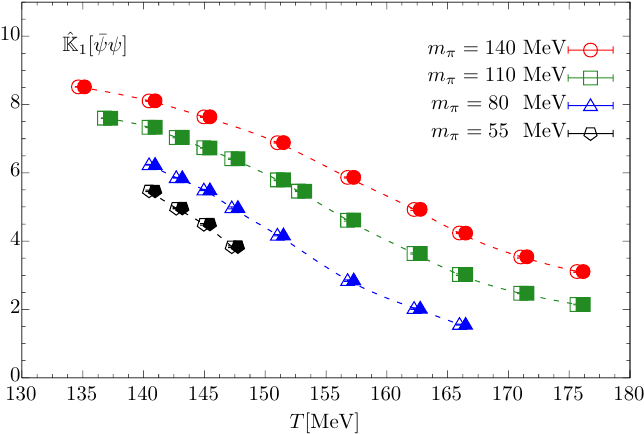}
    \includegraphics[width=0.32\textwidth,height=0.17\textheight]{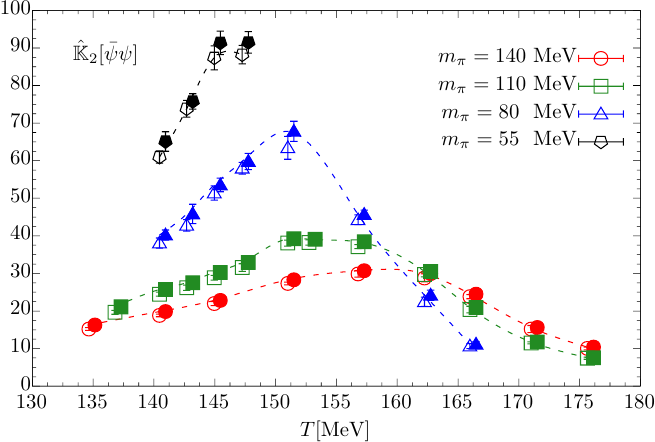}
    \includegraphics[width=0.32\textwidth,height=0.17\textheight]{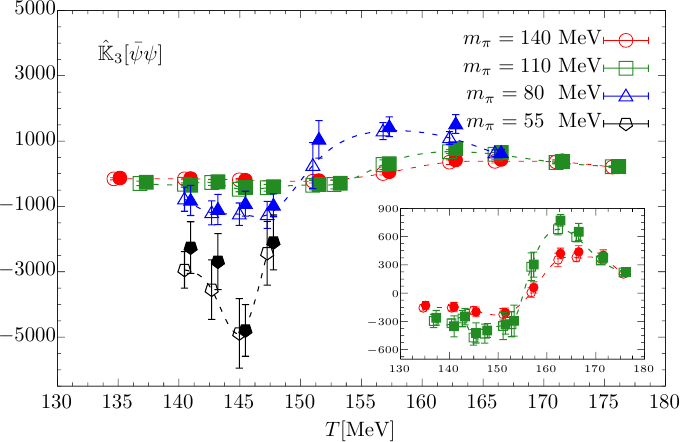}
  	\caption{Comparisons of direct measurements (open symbols) of $\cum{1}\left[ \pbp \right]$ (left), $\cum{2}\left[ \pbp \right]$ (middle) and $\cum{3}\left[ \pbp \right]$ (right) (\textit{cf.}~\autoref{eq:Kn-TrM}) with those reproduced (filled symbols, slightly shifted horizontally for visibility) from $P_n(\lambda)$ (\textit{cf.}~\autoref{eq:kn2}). The results are shown for all available values of light quark masses and temperatures. Lines which are just connections of data points and not fits are used to guide the eye.}
	\label{fig:K123-reproduce}
\end{figure}

\subsection{III.B. Supplemental results to \autoref{fig:P123} and \autoref{fig:RescaledP123}}

Supplementary to the results shown in \autoref{fig:P123} we show results of $\hp{n}$ with $n=1,2,3$ at higher temperatures ranging from 147~MeV to 176~MeV in \autoref{fig:P123-nonscaling}.

\begin{figure}[!htp]
  \centering
    \includegraphics[width=0.32\textwidth,height=0.17\textheight]{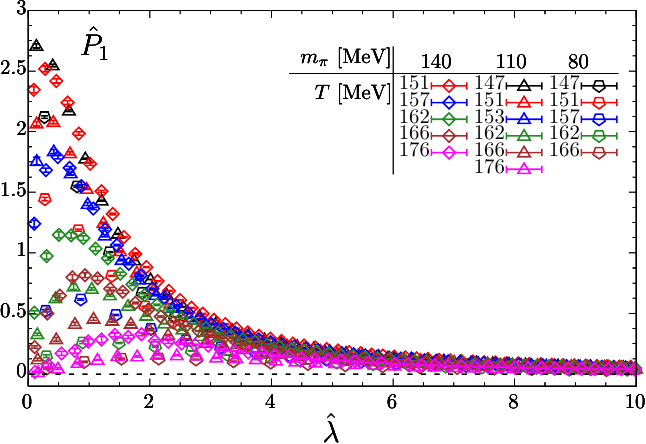}
    \includegraphics[width=0.32\textwidth,height=0.17\textheight]{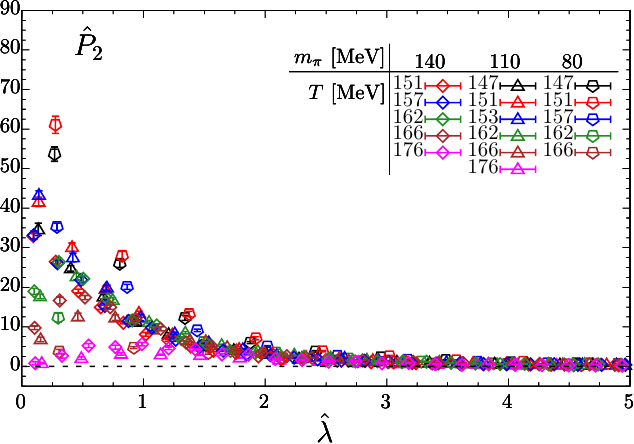}
    \includegraphics[width=0.32\textwidth,height=0.17\textheight]{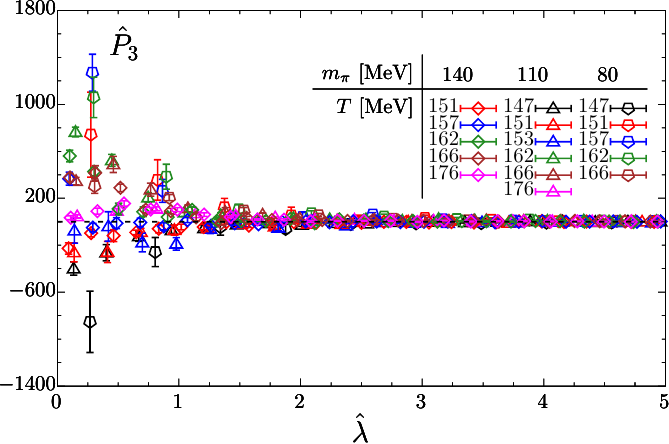}
  	\caption{Similar to~\autoref{fig:P123} but for 147~MeV~$\le T \le$~176~MeV, i.e. $\hp{1}$ (left), $\hp{2}$ (middle) and $\hp{3}$ (right).}
	\label{fig:P123-nonscaling}
\end{figure}

Supplementary to the results shown in \autoref{fig:RescaledP123}  we show results of $\hp{n}$ rescaled by $\hml^{1/\delta+1-n}f_n(z)$ for $n=1,2,3$ at higher temperatures ranging from 147~MeV to 176~MeV in \autoref{fig:RescaledP123-noscaling}.

\begin{figure}[!htp]
  \centering
    \includegraphics[width=0.32\textwidth,height=0.17\textheight]{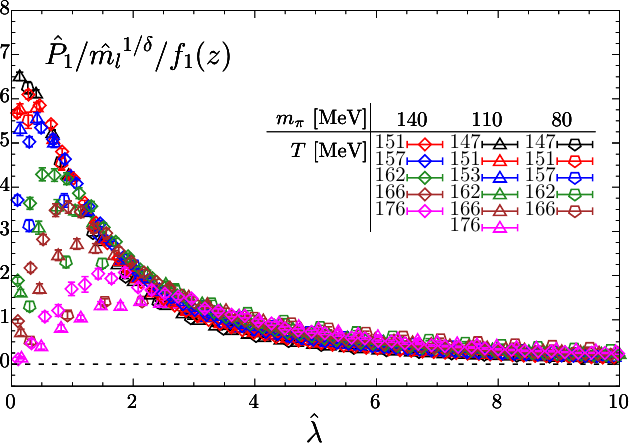}
    \includegraphics[width=0.32\textwidth,height=0.17\textheight]{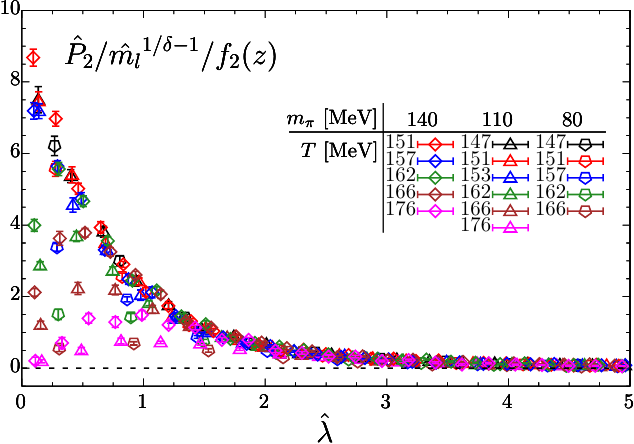}
    \includegraphics[width=0.32\textwidth,height=0.17\textheight]{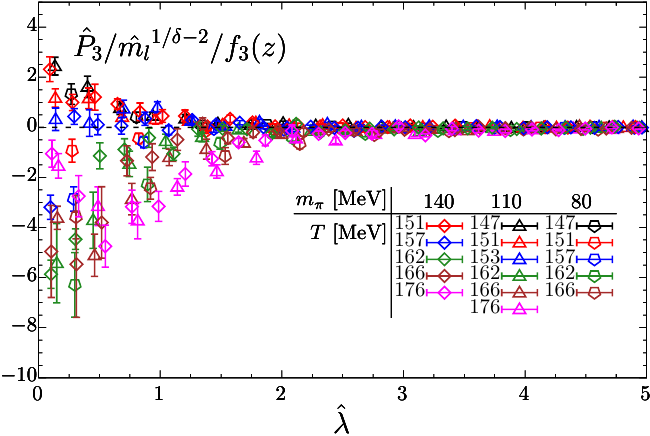}
  	\caption{Similar to~\autoref{fig:RescaledP123} but for 147~MeV~$\le T \le$~176~MeV, i.e. $\hp{n}$ in \autoref{fig:P123-nonscaling} rescaled by $\hml^{1/\delta+1-n}f_n(z)$ for $n=1$ (left), $n=2$ (middle) and $n=3$ (right).}
	\label{fig:RescaledP123-noscaling}
\end{figure}

Supplementary to the results shown in \autoref{fig:RescaledP123} we show $\hp{n}$ rescaled by $\hml^{1/\delta+1-n}f_n(z)$ for $n=1,2,3$ but using different values of $\{\tc,z_0\}$ to obtain $f_n(z)$ in \autoref{fig:ReacaledP123-z0=2.24-Tc=145.6}.  The values of $\{T_c,z_0\}$ adopted in \autoref{fig:ReacaledP123-z0=2.24-Tc=145.6} are $\tc=145.6(3)$ MeV and $z_0=2.24(5)$ as quoted in~\cite{Clarke:2020htu}, which are obtained from scaling fits to lattice data of static quark energy in a broader pion mass region of 55~MeV$\leq m_\pi\leq$140 MeV.  

\begin{figure}[!htp]
  \centering
    \includegraphics[width=0.32\textwidth,height=0.17\textheight]{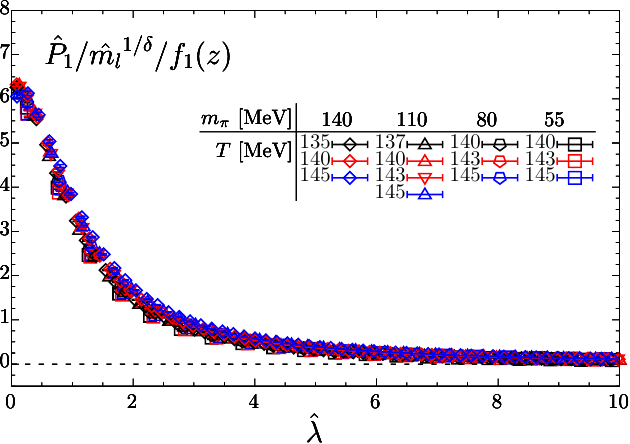}
    \includegraphics[width=0.32\textwidth,height=0.17\textheight]{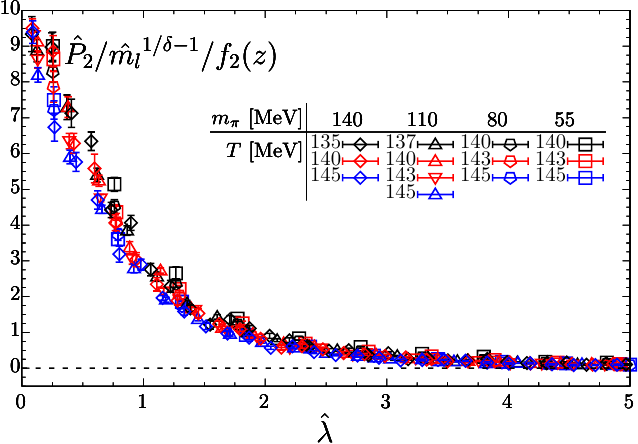}
    \includegraphics[width=0.32\textwidth,height=0.17\textheight]{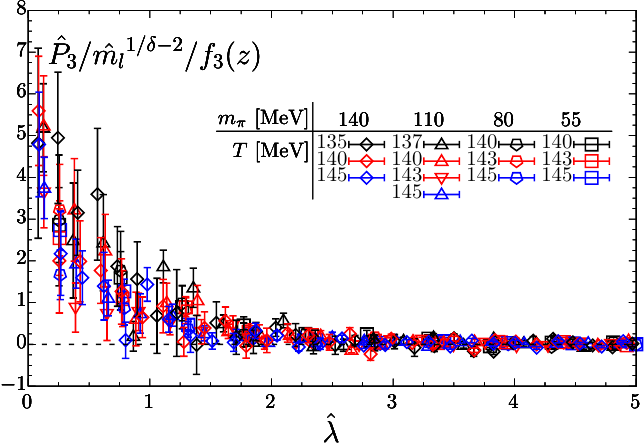}
  	\caption{Similar to~\autoref{fig:RescaledP123} but using $T_c=145.6(3)$ MeV and $z_0=2.24(5)$ as also quoted in~\cite{Clarke:2020htu} to obtain $f_n(z)$.}
	\label{fig:ReacaledP123-z0=2.24-Tc=145.6}
\end{figure}

\end{widetext}
\end{document}